\documentstyle[11pt,aaspp4,flushrt]{article}

\begin{document}

\vspace*{0.5in}
\title{PSR~J1518+4904: A Mildly Relativistic \\
  Binary Pulsar System}
\vspace*{0.1in}
\author{D. J. Nice}
\affil{National Radio Astronomy Observatory\altaffilmark{1}, \\
Edgemont Road, Charlottesville, VA 22903}
\author{R. W. Sayer \& J. H. Taylor}
\affil{Joseph Henry Laboratories and Physics Department, \\
Princeton University, Princeton, NJ 08544}
\vspace*{1in}
\centerline{Accepted for publication in {\it The Astrophysical Journal
(Letters)}.}
\altaffiltext{1}{Operated by Associated Universities, Inc., under
cooperative agreement with the National Science Foundation.}

\newpage

\begin{abstract}

PSR J1518+4904 is a recently discovered 40.9~ms pulsar in an 8.6~day,
moderately eccentric orbit.  We have measured pulse arrival times for
this pulsar over 1.4~yr at several radio frequencies, from which we
have derived high precision rotational, astrometric, and orbital
parameters.  The upper limit for the period derivative of the pulsar,
$\dot{P}<4\times 10^{-20}$, gives a characteristic age of at least
$1.6\times 10^{10}$\,yr, among the highest known.  We find the orbit to
be precessing at a rate of $0.0111\pm0.0002^\circ$\,yr$^{-1}$, which
yields a total system mass (pulsar plus companion) of
$2.62\pm0.07$\,M$_\odot$ according to general relativity.  
Further analysis of the orbital parameters
yields a firm upper limit of 1.75\,M$_\odot$ on the pulsar mass and
constrains the companion mass to the range 0.9 to 2.7\,M$_\odot$.
These masses, together with the sizable orbital eccentricity and other
evidence, strongly suggest that the companion is a second neutron star.

\end{abstract}


\newpage 

\section{Introduction}\label{sec:intro}

Fast-spinning ``recycled'' pulsars in binary systems are powerful
tools for the study of orbital kinematics.  Measurements of pulse
arrival times and the variation of time delays across a pulsar's orbit
provide a means of measuring orbital elements with very high
precision.  Further, pulsars and their orbiting companions are
generally compact enough that their motion can be treated as that of
two point masses.  Observations of binary pulsars have been used to
study the evolution of interacting binary systems, to measure
properties of neutron stars, and to test theories of gravitation
(\cite{tay92}).

Because of the high level of interest in such experiments, much effort
has been put into searching for further examples of relativistic
binary pulsars.  In this {\it Letter} we describe observations of
PSR~J1518+4904, a 41~ms pulsar in an eccentric, 8.6 day orbit with
what appears to be a second neutron star.  This pulsar was discovered
in a recent survey of 15,900\,deg$^2$ of the northern celestial
sphere.  The survey observations were made with the NRAO 140~foot telescope
at Green Bank, and data were processed with the Cray C90 of the
Pittsburgh Supercomputer Center.  The flux density limit was 8~mJy for
long-period pulsars, and the spectral sample interval was 256~$\mu$s.
In addition to PSR~J1518+4904, two other recycled pulsars were
detected: the relativistic binary PSR~B1534+12 (\cite{wol91a}) and the
white dwarf--neutron star binary PSR~J1022+1002 (\cite{cnst96}).
Details of the survey are given in Sayer, Nice \& Taylor
(1996)\nocite{snt96}.

\section{Observations}\label{sec:obs}
\begin{figure}
\plotone{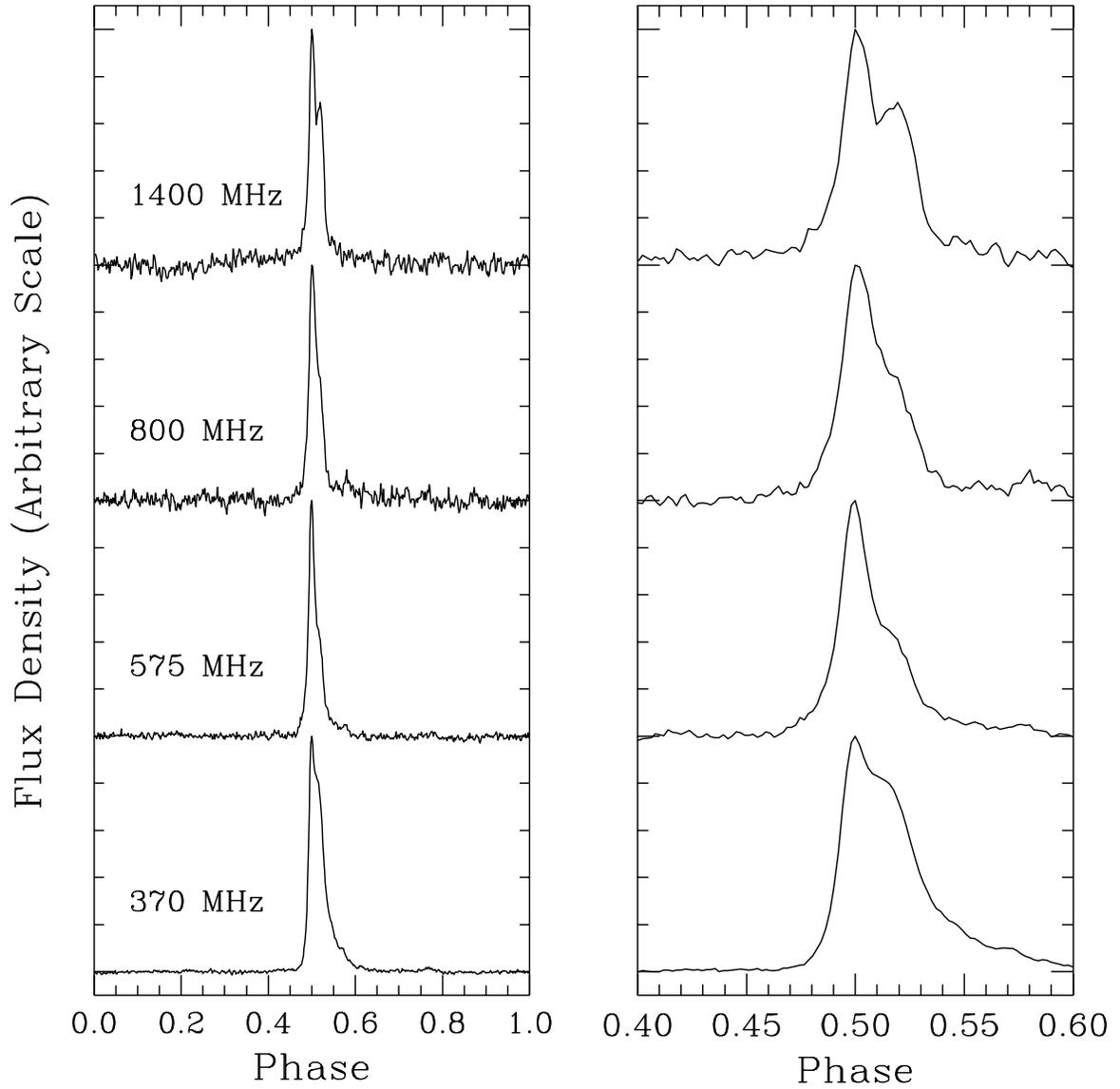}
\caption{\label{fig:prof}Pulse profile of PSR~J1518+4904 at several radio
frequencies.  {\it Left:} Full pulse period.  {\it Right:} Expanded view
of pulse peak.}
\end{figure}

We observed PSR~J1518+4904 with the 140~foot telescope on 83 days
between 28 October 1994 and 9 March 1995 at radio frequencies between
320 and 1400 MHz.  Most of the data were collected in approximately
bimonthly observations at 575 and 800 MHz, with observations typically
made on two days at each frequency; these observations were augmented
by two 10-day sessions at 370 MHz in March/April and
August/September 1995.  The extended sessions allowed full coverage of
all orbital phases at a single epoch.  Several shorter series of
observations were also made at 370 MHz, and brief observations were
made at 320 MHz and 1400 MHz.  Average pulse profiles at our principal
observing frequencies are presented in Figure~\ref{fig:prof}.

Data were collected with the Spectral Processor, a digital Fourier
transform spectrometer.  In each of two polarizations, 512 spectral
channels were synthesized across a 40 MHz passband.  The spectra were
folded synchronously at the predicted topocentric pulse period and
averaged for two minutes.  The resulting pulse profiles had 512 phase
bins, giving 80\,$\mu$s resolution.  Pulse arrival times were measured
in the conventional way: the data were de-dispersed after detection
and opposite polarizations summed to produce a single total-intensity
pulse profile for a given integration.  This profile was cross-correlated with
a standard template to measure the phase offset of the pulse within
the profile.  The offset was added to the start time and suitably
translated to the middle of the integration to yield an effective
pulse time of arrival.

Because the orbital ephemeris was not initially known, some early
observations were made with a ``search mode'' system in which spectra
were streamed to tape in raw form at intervals of 256\,$\mu$s.
Off-line processing used standard Fourier techniques to measure the
topocentric pulse period, and the spectra were then folded at that
period to generate profiles which were processed as described above.

We measured pulse times of arrival from 4496 profiles, excluding those
data in which the pulsar was not detected or in which the signal was
corrupted by radio frequency interference.  These arrival times were
averaged over spans of one hour, and a few low-quality points were
eliminated, yielding a final data set of 213 average time-of-arrival
measurements.  These were fit to a standard model of pulsar rotational
and orbital behavior using the program {\sc tempo} (\cite{tw89}).

\begin{figure}
\plotone{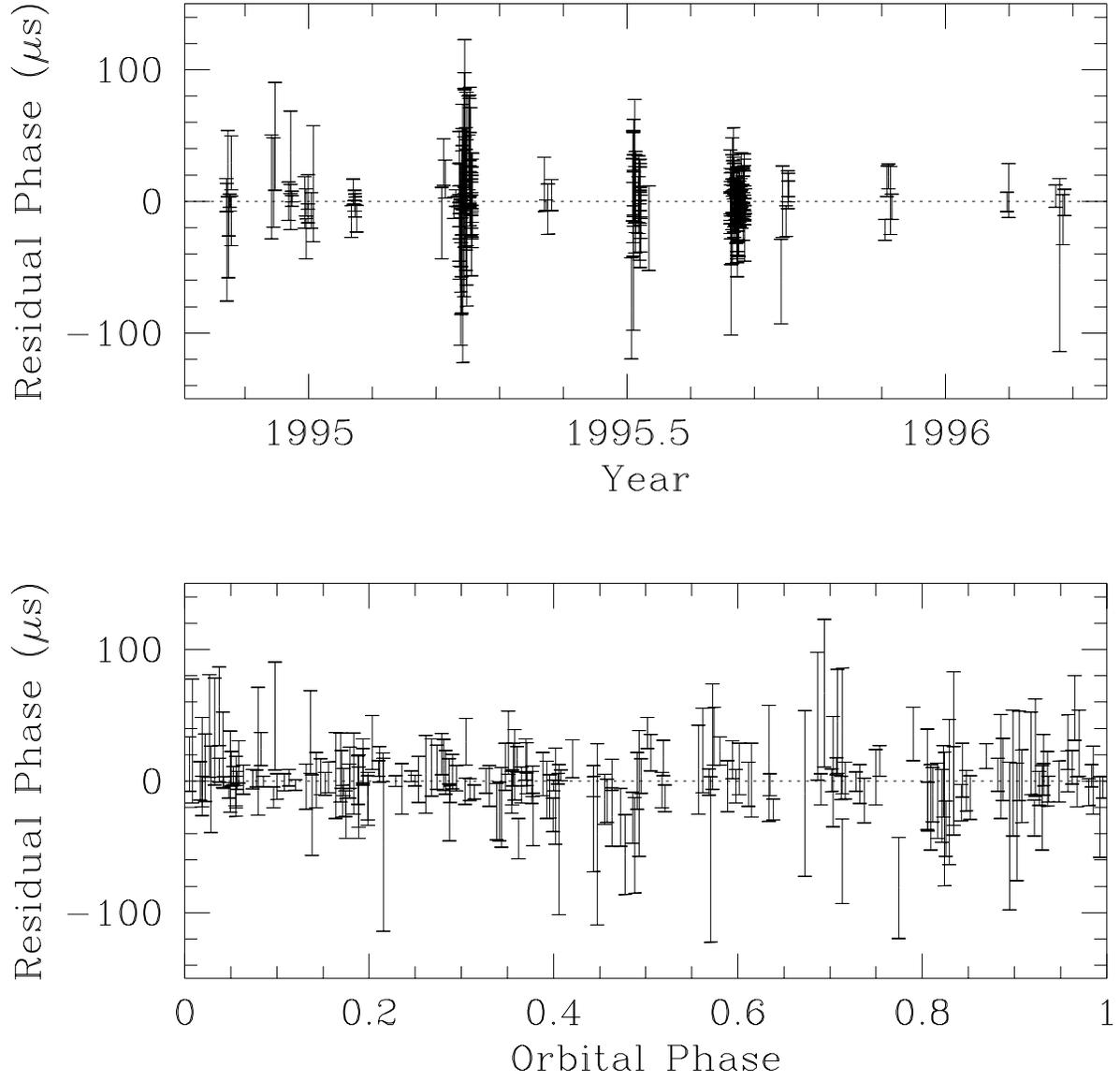}
\caption{\label{fig:res}Residual pulse arrival times after removing
the best-fit relativistic orbit model, plotted versus time and
orbital phase.}
\end{figure}

As expected for a tight, eccentric binary system, the five-parameter
Keplerian orbital model normally used to describe ``single-line
spectroscopic binary'' observations was not sufficient to remove fully
the orbital time-of-flight delays from the pulse arrival times.
However, we found that an excellent fit could be made by including
orbital precession in the timing model.  The measured rate of advance
of periastron is $\dot{\omega}=0.0111\pm0.0002^\circ$\,yr$^{-1}$.  The
full set of timing parameters is listed in Table~1.  
Uncertainties in the table are twice the formal errors of the
timing model;  given the high covariances between many of the
parameters, we believe these to be good estimates of the true 
uncertainties.  The
post-fit arrival times have root-mean-square residual of 20\,$\mu$s,
and appear to have random scatter (Figure~\ref{fig:res}).

\section{Analysis}\label{sec:results}

\subsection{Mass Measurements}

\begin{figure}
\plotone{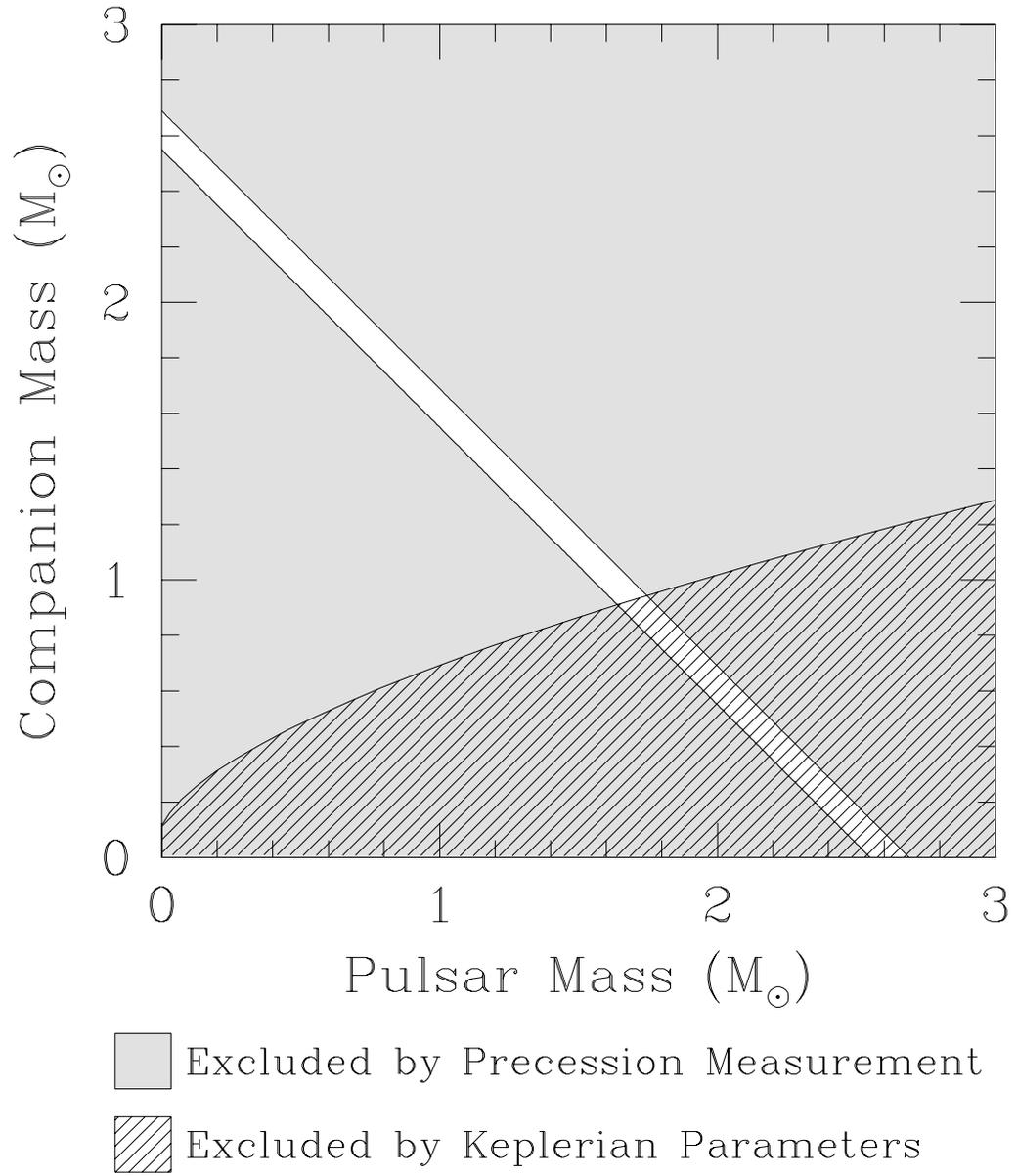}
\caption{\label{fig:mass}Constraints on the masses of the pulsar and
its companion.  The true values must lie in the open strip above
the hatched region.}
\end{figure}

Although the masses of the pulsar and companion cannot be
measured independently, the orbital elements constrain the allowed
values of these quantities.  First, the Keplerian ``mass function''
relates the masses of the pulsar, $m_1$, and its companion, $m_2$, to
the binary period $P_b$ and projected semi-major axis $a_1\sin i$
according to
\begin{equation}\label{eqn:massfunc}
\frac{(m_2\,\sin i)^3}{(m_1+m_2)^2} 
= \frac{(2\pi)^2}{T_\odot}\frac{(a_1\sin i)^3}{P_b^2},
\end{equation}
where $T_\odot=GM_\odot\,c^{-3}=4.925490947\times 10^{-6}$\,s,
$a_1\sin i$ is in light seconds, $P_b$ is in seconds,
and $m_1$ and $m_2$ are in solar
masses.  The orbital inclination $i$ is unknown; however, $\sin i$
must be less than 1, constraining the masses to lie outside the
hatched region at the lower-right of Figure~\ref{fig:mass}.

A second constraint on allowed masses comes from the precession of the
longitude of periastron.  Such precession can in principle arise from
tidal or rotational distortions of the companion, from the
relativistic gravitational interaction between the two bodies, or some
combination of these.  Given the size of the PSR~J1518+4904 orbit, and
assuming the companion is a compact object
(\S~\ref{sec:discussion}), the tidal and rotational distortion effects
can be neglected (\cite{sb76}), leaving only relativistic precession.
In this case $\dot{\omega}$ is a simple function of total system mass
$m_1+m_2$:
\begin{equation}
 \dot{\omega} = 3(2\pi/P_b)^{5/3}(1-e^2)^{-1} \,
  T_\odot^{2/3} \,(m_1+m_2)^{2/3}.
\end{equation}
The observed rate of precession, $\dot{\omega}=0.0111\pm
0.0002^\circ$\,yr$^{-1}$, yields total a system mass of
\begin{equation}
m_1+m_2=2.62\pm 0.07\,{\rm M}_\odot,
\end{equation}
constraining the component masses to lie in the narrow open strip outside
the gray region of Figure~\ref{fig:mass}.

While the masses of the pulsar and its companion cannot be fully
separated, Figure~\ref{fig:mass} shows that they must obey the
constraints $m_1<1.75$\,M$_\odot$ and $0.9<m_2<2.7$\,M$_\odot$,
respectively.  It is quite plausible that both $m_1$ and $m_2$ are
close to the Chandrasekhar mass, as has been observed in other
eccentric pulsar binaries (\S~\ref{sec:discussion}).  If the masses
are nearly equal, $m_1\approx m_2\approx1.31$\,M$_\odot$, then
Equation~\ref{eqn:massfunc} implies an orbital inclination angle 
$i=45^\circ$.

\subsection{Pulsar Spin-Down and Space Velocity}

Because our observations span little more than a year, the influences
of pulsar position, proper motion, and spin-down rate in the timing
solution are highly covariant, and we can presently
derive only upper limits for the period derivative and proper motion.
From the period derivative limit, $\dot{P}<4\times 10^{-20}$, we infer
the lower limit of the pulsar's characteristic age to be
$\tau=P/2\dot{P}>1.6\times 10^{10}$\,yr, among the highest measured.
While the true age of the pulsar is likely less than $1.6\times
10^{10}$\,yr, the high characteristic age suggests that the pulsar is
probably very old, and that it will evolve very little on a
Galactic timescale.  The period derivative limit constrains magnetic
field strength to be no more than $1.3\times 10^9$\,Gauss under the
conventional assumptions (\cite{mt77}).

The timing solution yields an upper limit of 60\,mas\,yr$^{-1}$ for the proper
motion of the pulsar.  This limit can be improved by noting that proper motion
$\mu$ of a pulsar at distance $d$ induces a pulse period derivative
$\dot{P}/P=\mu^2d/c$ (\cite{dt91}).  The observed limit $\dot{P}<4\times
10^{-20}$ and estimated distance of 0.7~kpc yield an upper limit of
30\,mas\,yr$^{-1}$.  This corresponds to a projected space velocity of no more
than 100\,km\,s$^{-1}$.  This is on the low end of the velocity distribution
of isolated pulsars (\cite{ll94}), but typical of recycled pulsars
(\cite{nt95}).

\section{Discussion}\label{sec:discussion}

There are several reasons to believe the pulsar's companion
is a neutron
star.  Optical observations detect nothing in the direction of the
pulsar to magnitude limits $m_B>24.5$ and $m_R>23$ (M. van
Kerkwijk, private communication), certainly ruling out a main sequence
companion.  The high eccentricity of the orbit is naturally explained
as a fossil of the supernova that formed the pulsar's companion.  Further,
it must be significant that four of the six probable
double-neutron-star binaries now known, including PSR J1518+4904, have
periods between 30 and 60\,ms (see Table~2).
Their measured total system masses are all similar, and whenever the
masses have been separable they are found to be nearly equal and close
to the Chandrasekhar limit.  We take these facts to be strong
circumstantial evidence that the companions are neutron stars.

Double neutron star binaries are generally believed to evolve from
high mass X-ray binaries after a stage of common-envelope evolution
and spiral-in (e.g., \cite{ver93}).  However, van den Heuvel, Kaper \&
Ruymaekers (1994)\nocite{vkr94} argue that it is difficult to produce
the wide orbit of PSR~B2303+46 in this way because of the
extremely wide pre-spiral-in orbit that is required.  They suggest the
second neutron star in this system must have evolved from a very
massive ($>40-45$\,M$_\odot$) star which expelled its envelope without
extracting energy from the orbital system (see also Kaper {\it et al.}
1995\nocite{klr+95}).  Given that the PSR~J1518+4904 orbit is similar
in size to that of PSR~B2303+46, we suppose that the PSR~J1518+4904
system must have evolved in a similar way.

Further evidence for this scenario comes from the large characteristic age
of PSR J1518+4904, which implies that the pulsar's present-day
period and period derivative are close to those of the pulsar immediately
after spin-up.  This contrasts with PSRs~B1534+12, B1913+16, and B2127+11C, in
which the observed periods and period derivatives place the pulsars close to
the ``spin-up line'', the period--magnetic field relation expected after
Eddington accretion onto a magnetized neutron star (e.g., \cite{bv91}).  The
parameters of PSR~J1518+4904 place it far from the spin-up line.  A simple
explanation for this is that accretion onto the neutron star was far below the
Eddington rate (\cite{ctk94}), which agrees well with the model in which the
companion's stellar envelope was largely shed of its own accord rather than
through binary spiral-in.

Because the J1518+4904 system is much wider than the prototypical
double neutron star system of PSR~B1913+16, it emits far less
gravitational radiation, and the relativistic decay of its orbit will
be difficult to detect.  Using the formulae of Peters
(1964)\nocite{pet64}, and assuming that the pulsar and its companion
have equal masses, we calculate the coalescence time of the J1518+4904
system due to gravitational radiation to be $2.4\times 10^{12}$\,
years.  Inspiraling double neutron star binaries are a prime target
for gravitational observatories such as LIGO, so the birth rates and
lifetimes of these systems are of great interest (\cite{phi91}, \cite{cl95}).
For obvious reasons, the discovery of PSR~J1518+4904 does not
significantly influence such merger-rate calculations.

\acknowledgements

We thank Z.~Arzoumanian for assisting with the timing observations of
this pulsar.  The 140\,foot telescope of the National Radio Astronomy
Observatory is a facility of the National Science Foundation, operated
by Associated Universities, Inc., under a cooperative agreement.  This
pulsar was discovered in data analysis performed under grant
AST930018P from the Pittsburgh Supercomputer Center, sponsored by the
NSF.  Pulsar research at Princeton University is sponsored by NSF
grant AST~91-15103.

\clearpage

\begin{table}
\begin{center}

% \caption{\label{tab:param}Parameters of PSR J1518+4904}
\centerline{Table 1.  Parameters of PSR J1518+4904$^a$}
\vspace*{5mm}

\begin{tabular}{ll}
\hline\hline
\multicolumn{2}{c}{Measured Parameters}  \\
\hline
Period (ms)                          \dotfill & 40.93498826871(6)           \\
Period Derivative                    \dotfill & $<4\times10^{-20}$          \\
Right Ascension (J2000)        \dotfill & $15^{\rm h}18^{\rm m}16\fs799(1)$ \\
Declination (J2000)                  \dotfill & $+49^\circ04'34\farcs29(2)$ \\
Proper Motion (mas yr$^{-1}$)        \dotfill & $<60^b$                     \\
Epoch (MJD)                          \dotfill & 49894.00                    \\
Dispersion Measure (pc cm$^{-3}$)    \dotfill & 11.611(5)                   \\
Orbital Period (days)                \dotfill &      8.63400485(15)          \\
Projected Semi-Major Axis (light sec)\dotfill &      20.044003(4)           \\
Eccentricity                         \dotfill &      0.2494849(3)           \\
Angle of Periastron                  \dotfill &      342\fdg46217(8)        \\
Time of Periastron (MJD)             \dotfill &      49896.246989(2)        \\
Rate of Advance of Periastron ($^\circ$\,yr$^{-1}$)\ldots  &  0.0111(2)     \\
\hline
\multicolumn{2}{c}{Derived Parameters}  \\
\hline
Galactic Latitude                    \dotfill &   $54\fdg3$                 \\
Galactic Longitude                   \dotfill &   $80\fdg$8                 \\
Distance (kpc)                       \dotfill &   $0.70^{+0.13}_{-0.07}$     \\
Magnetic Field (G)                   \dotfill &   $< 1.3\times 10^{9}$      \\
Age (yr)                             \dotfill &   $> 1.6\times 10^{10}$     \\
Total System Mass (M$_\odot$)        \dotfill & $2.62(7)$                   \\
Orbit Decay Time (yr)                \dotfill &   $ 2400 \times 10^{9}$     \\
\hline
\end{tabular}
\end{center}
$^a$Figures in parenthesis are uncertainties in the last digit quoted. \\
$^b$A limit of $\mu<30$\,mas\,yr$^{-1}$ can be derived from the period
    derivative limit;  see text.
\end{table}

\clearpage

\begin{table}
%\caption{\label{tab:nsns}Neutron Star--Neutron Star Binaries}
\centerline{Table 2.  Neutron Star--Neutron Star Binaries $^{a,b}$ }
\vspace*{5mm}

\begin{tabular}{lrrrrclll}
\hline\hline
\multicolumn{1}{c}{PSR}
& \multicolumn{1}{c}{$P$} & \multicolumn{1}{c}{$P_b$} & $a_1\,\sin i$ 
& \multicolumn{1}{c}{$e$}
& \multicolumn{1}{c}{$\dot{\omega}$} & \multicolumn{1}{c}{$m_1+m_2$}     
& \multicolumn{1}{c}{$m_1$}     & \multicolumn{1}{c}{$m_2$} \\
&  \multicolumn{1}{c}{(ms)}         & (days)& (lt-sec)   
&     & ($^\circ$\,yr$^{-1}$)
&  \multicolumn{1}{c}{(M$_\odot$)} 
& \multicolumn{1}{c}{(M$_\odot$)} & \multicolumn{1}{c}{(M$_\odot$)} \\
\hline
J1518+4904&   40.93  & 8.634 &  20.04    & 0.249 & 0.011 & 2.66(7) & & \\
B1913+16  &   59.03  & 0.323 &   2.34    & 0.617 & 4.227 & 2.8284  & 
1.44 &  1.39      \\
B1534+12  &   37.90  & 0.420 &   3.73    & 0.274 & 1.756 & 2.6784  & 
1.34 &  1.34      \\
B2127+11C &   30.53  & 0.335 &   2.52    & 0.681 & 4.457 & 2.712   & 
1.35(4) & 1.36(4)   \\
B2303+46  & 1066.37  &12.340 &  32.69    & 0.658 & 0.010 & 2.60(6) &  &     \\ 
B1820$-$11$^c$ & 279.83 &357.762& 200.67 & 0.795 & $\lesssim\!10^{-4}$ & & & \\
\hline\hline
\end{tabular}
\vspace*{5mm}

$^a$Figures in parenthesis are uncertainties in the last digit quoted.  Where no
such figure is given uncertainty is $\pm1$ or less in the last digit quoted.

$^b$References: \cite{arz95} (B1534+12, B1820$-$11, B2303+46); \cite{dk96}
(B2127+11C); \cite{tw89} (B1913+16); \cite{tamt93} (B2303+46); \cite{wol91a}
(B1534+12).

$^c$May not be a neutron star--neutron star system

\end{table}

\clearpage

\clearpage

\clearpage

\end{document}